\documentclass[a4paper,12pt, epsfig]{article}
\def\letter{0}\def\pr{0}
\pdfoutput=1 %per jhep
\usepackage{epsfig}
\usepackage{epstopdf}
\usepackage{graphicx}
\usepackage{ifthen}
\usepackage{ulem}

\usepackage{amsmath}
\usepackage[psamsfonts]{amssymb}
\usepackage{euscript}

\usepackage{latexsym}
\usepackage[arrow,matrix,curve]{xy}

\usepackage[hypertexnames=false]{hyperref}
\hypersetup{
    colorlinks=true,
    linkcolor=blue,
    filecolor=magenta,   
    citecolor=black,   
    urlcolor=cyan,
    }

\newskip\humongous \humongous=0pt plus 1000pt minus 1000pt

\newif\ifdtup

\def\,{\hspace{-.1cm}}
\def\hsp{,\hspace{.7cm}}

\def\fc#1#2 {\frac{n}{q}#1\frac{n}{q}#2}

\def\kt{\kappa}
\def\kt{\mathfrak{K}}
\def\ks{|\kt\rangle}

\def\rv{{\rm{vac}}}

\newcommand{\vac}{\ensuremath{|0\rangle}}

\renewcommand{\sin}{\textrm{sin}}

\renewcommand{\sinh}{\textrm{sinh}}

\renewcommand{\tanh}{\textrm{tanh}}
\newcommand{\sech}{\textrm{sech}}
\newcommand{\csch}{\textrm{csch}}

\def\sign#1{{\rm sign}\left(#1\right)}

\renewcommand{\theequation}{\arabic{section}.\arabic{equation}}
\renewcommand{\(}{\begin{equation}}
\renewcommand{\)}{end{equation} \vspace{-.05in}\linebreak}

\newcounter{saveeqn}
\newcounter{savealpheqn}

\newcommand{\alpheqn}{\setcounter{saveeqn}{\value{equation}}%
  \stepcounter{saveeqn}\setcounter{equation}{0}%
  \renewcommand{\theequation}{\mbox{\arabic{section}.\arabic{saveeqn}
\alph{equation}}}
  \renewcommand{\)}{\end{equation}}}
\def\part#1{\frac{\partial}{\partial{#1}}}%
\def\group#1{\refstepcounter{equation}\setcounter{saveeqn}
 {\value{equation}}%
  \label{#1}\setcounter{equation}{0}%
\renewcommand{\theequation}{\mbox{\arabic{section}.\arabic{saveeqn}
\alph{equation}}}
  \renewcommand{\)}{\end{equation}}}
\newcommand{\reseteqn}{\setcounter{equation}{\value{saveeqn}}%
  \renewcommand{\theequation}{\arabic{section}.\arabic{equation}}%
  \renewcommand{\)}{\end{equation}}}

\newcommand{\aalpheqn}{\setcounter{saveeqn}{\value{equation}}%
  \stepcounter{saveeqn}\setcounter{equation}{0}%
  \renewcommand{\theequation}{\mbox{
        \Alph{subsection}.\arabic{saveeqn}\alph{equation}}}
   \renewcommand{\)}{\end{equation}}}
\newcommand{\areseteqn}{\setcounter{equation}{\value{saveeqn}}%
  \renewcommand{\theequation}{\Alph{subsection}.\arabic{equation}}%
  \renewcommand{\)}{\end{equation}}}

\renewcommand{\thefootnote}{\alph{footnote}}
\renewcommand{\(}{\begin{equation}}
\renewcommand{\)}{\end{equation}}
\newcommand{\ba}{\begin{eqnarray}}
\newcommand{\ea}{\end{eqnarray}}

\renewcommand{\b}{\beta}

\renewcommand{\sl}{{\sqrt{\lambda}}}

\newcommand{\cbp}{\mathop{\vtop{\ialign{##\crcr
   $\hfil\displaystyle{}\hfil$\crcr\noalign{\kern-13pt\nointerlineskip}
   \BIG{)}\hskip0pt\crcr\noalign{\kern3pt}}}}}
\newcommand{\pa}{\mathop{\vtop{\ialign{##\crcr

$\hfil\displaystyle{\oplus}\hfil$\crcr\noalign{\kern+1pt\nointerlineskip
}
   \hspace{.08in}$^{\alpha=0}$\hskip6pt\crcr\noalign{\kern3pt}}}}}
\renewcommand{\hsp}{,\hspace{.3in}}
\newcommand{\p}{^\prime}

\catcode`\@=11
\def\vereq#1#2{\lower3pt\vbox{\baselineskip1.5pt \lineskip1.5pt
\ialign{$\m@th#1\hfill##\hfil$\crcr#2\crcr\sim\crcr}}}
\catcode`\@=12

\renewcommand{\(}{\begin{equation}}
\renewcommand{\)}{\end{equation}}

\def\pin#1{\int \frac{d#1}{2\pi}}
\def\ppin#1{\int\hspace{-17pt}\sum \frac{d#1}{2\pi}}
\def\ppink#1{\int\hspace{-17pt}\sum\frac{d^{#1}k}{(2\pi)^{#1}}}
\def\ppinkp#1{\int\hspace{-17pt}\sum\frac{d^{#1}k\p}{(2\pi)^{#1}}}
\def\dint{\int\hspace{-12pt}\sum }
\def\pink#1{\int \frac{d^{#1}k}{(2\pi)^{#1}}}

\def\pinkp#1{\int \frac{d^{#1}k\p}{(2\pi)^{#1}}}

\def\Bd#1{B^\ddag_{k_{#1}}}

\def\cc{\mathcal{C}}
\def\df{\mathcal{D}_{f}}

\def\I{\mathcal{I}}

\def\os{\omega_S}

\def\blu#1{\textcolor{blue}{Jarah: #1}}
\def\red#1{\textcolor{red}{Hui: #1}}

\newcommand{\beas}{\begin{eqnarray*}}
\newcommand{\eeas}{\end{eqnarray*}}

\newcommand{\bquo}{\begin{quote}}
\newcommand{\enqu}{\end{quote}}

\def\lim#1{\stackrel{\rm{lim}}{{}_{#1}}}

\newcommand{\R}{{\mathbb R}}

\newcommand{\g}{\mathfrak g}

\def\ok#1{\omega_{k_{#1}}}
\def\okp#1{\omega_{k\p_{#1}}}
\def\okt#1{\omega_{\kt_{#1}}}
\def\V#1{V^{(#1)}(\sqrt{\lambda}f(x))}
\def\v#1{V^{(#1)}[f(x),x]}
\def\ck{\csch\left(\frac{\pi k}{2\b}\right)}

\def\mb{\mathcal{B}}
\def\mc{\mathcal{C}}
\def\md{\mathcal{D}}
\def\me{\mathcal{E}}

\newcommand{\beq}{\begin{equation}}
\newcommand{\eeq}{\end{equation}}
\newcommand{\bea}{\begin{eqnarray}}
\newcommand{\eea}{\end{eqnarray}}

\newskip\humongous \humongous=0pt plus 1000pt minus 1000pt

\newif\ifdtup

\catcode`\@=11

\ifthenelse{\equal{\letter}{0}}{ %se lettere=0, commenta questo

\@addtoreset{equation}{section}
\def\theequation{\arabic{section}.\arabic{equation}}

\def\@normalsize{\@setsize\normalsize{15pt}\xiipt\@xiipt
\abovedisplayskip 14pt plus3pt minus3pt%
\belowdisplayskip \abovedisplayskip
\abovedisplayshortskip \z@ plus3pt%
\belowdisplayshortskip 7pt plus3.5pt minus0pt}

\def\small{\@setsize\small{13.6pt}\xipt\@xipt
\abovedisplayskip 13pt plus3pt minus3pt%
\belowdisplayskip \abovedisplayskip
\abovedisplayshortskip \z@ plus3pt%
\belowdisplayshortskip 7pt plus3.5pt minus0pt
\def\@listi{\parsep 4.5pt plus 2pt minus 1pt
      \itemsep \parsep
      \topsep 9pt plus 3pt minus 3pt}}

\relax

\def\section{\@startsection{section}{1}{\z@}{3.5ex plus 1ex minus  .2ex}{2.3ex plus .2ex}{\large\bf}}

\def\thesection{\arabic{section}}
\def\thesubsection{\arabic{section}.\arabic{subsection}}

\def\appendix{\setcounter{section}{0}
 \def\thesection{Appendix \Alph{section}}
 \def\thesubsection{\Alph{section}.\arabic{subsection}}
 \def\theequation{\Alph{section}.\arabic{equation}}}
\renewcommand{\theequation}{\arabic{section}.\arabic{equation}}

}{
\renewcommand{\theequation}{\arabic{equation}}

} %se letter=0, commenta questo

\begin{document}
% ========================================================================
\def\thefootnote{\fnsymbol{footnote}}
\def\thetitle{Elastic Kink-Meson Scattering}
\def\auttwo{Hui Liu}
\def\autone{Jarah Evslin}
\def\affc{Institute of Physics, Polish Academy of Sciences, Aleja Lotników 32/46, 02-668 Warsaw, Poland}
\def\affb{University of the Chinese Academy of Sciences, YuQuanLu 19A, Beijing 100049, China}
\def\affa{Institute of Modern Physics, NanChangLu 509, Lanzhou 730000, China}
%\def\affe{Institute of Contemporary Mathematics, School of Mathematics and Statistics,Henan University, Kaifeng, Henan 475004, P. R. China}
%\def\affc{Departamento de F\'isica de Part\'iculas, Universidad deSantiago de Compostela and Instituto Galego de F\'isica de Altas Enerxias% (IGFAE), E-15782
%Santiago de Compostela, Spain}
%\def\affd{Physics Dept., Brookhaven National Laboratory, Bldg. 510A, Upton, NY 11973, USA}
%\def\affc{Wigner Research Centre, H-1525 Budapest 114, P.O.B. 49, Hungary}

%\title{Titolo}

\ifthenelse{\equal{\pr}{1}}{
\title{\thetitle}
\author{\autone}
\author{\auttwo}
\author{\autthree}
\affiliation {\affa}
\affiliation {\affb}
%\affiliation {\affc}
%\affiliation {\affd}
%pr e uno

}{

\begin{center}
{\large {\bf \thetitle}}

\bigskip

\bigskip

%\catcode`@=11

{\large \noindent  \autone{${}^{1,2}$} \footnote{jarah@impcas.ac.cn} %{
and \auttwo{${}^{3}$} \footnote{hliu@ifpan.edu.pl}
%and \autthree{${}^{5}$} \footnote{byzhang@henu.edu.cn}
}

%{\large \noindent  \autone{${}^{1,2}$} \footnote{jarah@impcas.ac.cn} and \auttwo{${}^{1,2}$} \footnote{guohengyuan@impcas.ac.cn}}

\vskip.7cm

1) \affa\\
2) \affb\\
3) \affc\\
%4) \affd\\
%5) \affe\\

\end{center}

}

\begin{abstract}
\noindent
In classical field theory, radiation does not reflect off of reflectionless kinks.  In quantum field theory, radiation quanta, called mesons, can be reflected.  We provide a general analytical formula for the leading order amplitude and probability for the elastic scattering of mesons off of reflectionless quantum kinks.  In the case of the Sine-Gordon model we verify that, due to a cancellation of six contributing processes, our general formula yields an amplitude of zero, as is required by integrability.

\end{abstract}

% \vfill
%
% \end{titlepage}
\setcounter{footnote}{0}
\renewcommand{\thefootnote}{\arabic{footnote}}

\ifthenelse{\equal{\pr}{1}}
{
\maketitle
}{}

\section{Introduction}

Just 40 years ago, the scattering of quantum solitons with fundamental quanta was a popular topic \cite{weigel89}.  The strategy was as follows.  First, one would consider kinks in (1+1)-dimensional models, using the collective coordinate approach of Refs.~\cite{gs74,tom75}.  For example, the elastic meson-kink scattering considered in the present work was first treated using collective coordinates in Ref.~\cite{uehara91}.  Once the lower-dimensional case was understood, the results would be imported into higher dimensional models \cite{hayashi92,erase}.  Suitable approximations were made \cite{adiabatic84} in this formal work and it was then applied to phenomenology \cite{diakonov97,ellis04}.

The house of cards reached its peak with the discovery of a predicted exotic hadron in Ref.~\cite{nakano03}.  Within a few years it became clear \cite{notheta} that this resonance, whose prediction was reasonably independent of the details of the model \cite{petrov16}, did not exist.  In fact, many of the predictions made over the previous twenty years were falsified one at a time.  Of course the problem could be with assumptions built into the models, or, as advocated in Ref.~\cite{weigel18}, it could be with the approximations used to treat calculations involving quantum solitons\footnote{Indeed, the importance of including the full continuum quantum degrees of freedom has recently by highlighted in Refs.~\cite{chris23,bjarke23}.}.

This second possibility motivates a lighter formalism, so that less severe approximations are necessary and as a result more reliable conclusions may be expected.  Such a formalism, linearized soliton perturbation theory, has been formulated in Refs.~\cite{mekink,me2loop}.  

The purpose of the present paper is to re-examine elastic meson-kink scattering using this new, simpler formalism.  Our answer, summarized in Eqs.~(\ref{abcd}) and (\ref{peq}), will disagree with the old result, Eq. (3.19) of Ref.~\cite{uehara91}.  At leading order, we find, for example, that there are contributions from states with two virtual mesons.  As we will show, such contributions in fact are essential to eliminate soliton-meson elastic scattering in the Sine-Gordon model, which, as their masses differ, would be in contradiction with the integrability of the model.  That contribution is not present in Ref.~\cite{uehara91} as loop contributions were intentionally dropped.  However that calculation, like ours, kept contributions to the amplitude that are quadratic in the coupling constant, which are of the same order as these loop corrections.  Indeed, this is evidenced by the fact that the loop corrections cancel the other contributions in the Sine-Gordon case.

Besides the fact that they are essential for consistency, these intermediate two-meson states are interesting for another reason.  In models in which the kink has bound normal mode excitations, called shape modes, there will be an intermediate state consisting of a twice-excited shape mode.  In models such as the $\phi^4$ model,  a single shape mode excitation is stable while a double excitation is unstable.  We therefore expect that our scattering probability will have a narrow peak at twice the energy of the shape mode, with a width equal to the shape mode's inverse lifetime.  More generally, we hope that kink-meson scattering can teach us about the unstable excited spectrum of the kink itself.  We will test this general expectation in future work, but the aforementioned peak will already be visible below.

We begin in Sec.~\ref{revsez} with a review of linearized soliton perturbation theory.  Our main calculation is presented in Sec.~\ref{calcsez}.  Finally, we turn to the Sine-Gordon model in Sec.~\ref{exsez}.

\section{Linearized Soliton Perturbation Theory} \label{revsez}

Consider a  (1+1)-dimensional quantum field theory with a scalar field $\phi(x)$ and its conjugate field $\pi(x)$.  We will work in the Schrodinger picture, where the Hamiltonian may be written as
\begin{equation}
H=\int d x: \mathcal{H}(x):_a\hsp \mathcal{H}(x)=\frac{\pi^2(x)}{2}+\frac{\left(\partial_x \phi(x)\right)^2}{2}+\frac{V(\sqrt{\lambda} \phi(x))}{\lambda}
\end{equation}
in terms of a degenerate potential $V$ and an expansion parameter $\sqrt{\lambda}$.  The corresponding classical equations of motion enjoy a stationary kink solution $\phi(x,t)=f(x)$ that interpolates between the minima of the potential.  

The usual normal ordering $::_a$ removes ultraviolet divergences arising from loops connected to a single vertex, which are the only ultraviolet divergences in such theories.  The normal ordering is defined at the mass scale $m$ where
\beq
m^2=V^{(2)}(\sqrt{\lambda} f(\pm \infty))\hsp
V^{(n)}(\sqrt{\lambda} \phi(x))=\frac{\partial^n V(\sqrt{\lambda} \phi(x))}{(\partial \sqrt{\lambda} \phi(x))^n}.
\eeq
If the masses corresponding to the two sign choices are different, then at one loop the kink will accelerate \cite{wstabile}.  We will not be interested in such cases.

We refer to the quanta of the $\phi(x)$ field as mesons.  The collection of states with no kinks, but a finite number of mesons, will be called the vacuum sector.  States with a single kink, together with a finite number of mesons, are referred to as the kink sector.  Any kink sector state may be created by acting the displacement operator
\beq
\df={{\rm Exp}}\left[-i\int dx f(x)\pi(x)\right]\hsp
\df^\dag \phi(x) \df = \phi(x)+f(x)  \label{dfd}
\eeq
on a vacuum sector state.  Intuitively this is clear, as $\df$ shifts the field $\phi(x)$ by the classical kink solution.

It may seem that the problem of constructing kink sector states is nonperturbative, as the operator $\df$ contains an exponential of $f(x)$ which is proportional to $1/\sl$.  This apparent problem can be resolved using a passive transformation to remove the $\df$ from each state.  More specifically, first we choose names $|\psi\rangle$ for all of the states.  This choice of names for kets is called the defining frame.  We will work in a different frame, called the kink frame, defined as follows.  In the kink frame, the state $|\psi\rangle$ is defined to be the state $\df^\dag|\psi\rangle$ in the defining frame.  One can easily show that if this state is time-independent, so that $\df^\dag|\psi\rangle$ is an eigenstate of $H$, then $|\psi\rangle$ is an eigenstate of the kink Hamiltonian
\beq
H\p=\df^\dag H\df%\hsp P\p=\df^\dag P\df=P+\sqrt{Q_0}\pi_0
. 
\label{df}
\eeq
This is a manifestation of the familiar fact that, when making a passive transformation of coordinates, in this case coordinates $|\psi\rangle$ on the Hilbert space, one must remember to also transform all of the functions that act on those coordinates, in this case the operators.

What have we gained with this passive transformation?  Now all of the $\df$ operators have been removed from the names of our kink sector states, thus we can treat them using ordinary perturbation theory, with the caveat that the Hamiltonian in this frame is $H\p$.

In the vacuum sector, perturbation theory describes small perturbations about the vacuum.  Classically the constant frequency perturbations are plane waves.  In the kink sector, kink frame perturbation theory describes small perturbations about the kink.  Classically, small constant frequency $\omega$ perturbations are normal modes
\beq
\phi(x,t)=e^{-i\omega t}\g(x)
\eeq
which satisfy the Sturm-Liouville equation
\beq
\V{2}{\g}(x)=\omega^2{\g}(x)+{\g}^{\prime\prime}(x).  \label{sl}
\eeq
The solutions are classified by their frequencies $\omega$.  There is a single zero-mode $\g_B(x)$ with $\omega_B=0$.   For each real number $k$ there is a continuum mode $\g_k(x)$ with frequency
\beq
\ok{}=\sqrt{m^2+k^2}.
\eeq
There can also be discrete, real shape modes $\g_S(x)$ with frequencies $0<\omega_S<m$.   All modes are chosen to satisfy $\g^*_k=\g_{-k}$ and the completeness relations
\beq
\int dx |{\g}_{B}(x)|^2=1,\
\int dx {\g}_{k_1} (x) {\g}^*_{k_2}(x)=2\pi \delta(k_1-k_2),\ 
\int dx {\g}_{S_1}(x){\g}^*_{S_2}(x)=\delta_{S_1S_2}. \label{comp}
\eeq
The sign of $\g_B$ is fixed by the convention
\beq
\g_B(x)=-\frac{f\p(x)}{\sqrt{Q_0}}. \label{gb}
\eeq
Here $Q_i$ is the $O(\lambda^{i-1})$ term in the mass of the ground state kink.

Following Refs.~\cite{cahill76,mekink} we decompose the field and its conjugate as
\bea
\phi(x) &=&\phi_0 \mathfrak{g}_B(x)+\ppin{k} \left(B_k^{\ddag}+\frac{B_{-k}}{2 \omega_k}\right) \mathfrak{g}_k(x)\hsp
B^\ddag_k=\frac{B^\dag_k}{2\ok{}}\hsp
B^\ddag_S=\frac{B^\dag_S}{2\omega_S} \label{dec}\\
\pi(x) &=&\pi_0 \mathfrak{g}_B(x)+i \ppin{k}\left(\omega_k B_k^{\ddag}-\frac{B_{-k}}{2}\right) \mathfrak{g}_k(x)\hsp
B_S=B_{-S}\hsp \ppin{k}=\pin{k}+\sum_S. \nonumber
\eea
Here $\phi_0$ and $\pi_0$ represent the position and momentum of the kink center of mass.  The operators $B^\ddag_S$ and $B_S$ create and annihilate shape modes, while $B^\ddag_k$ and $B_k$ create and annihilate continuum modes.  The canonical commutation relations satisfied by $\phi(x)$ and $\pi(x)$ yield the commutators of $\pi_0,\ \phi_0, B^\ddag$ and $B$
\beq
\left[\phi_0, \pi_0\right]=i, \quad\left[B_{S_1}, B_{S_2}^{\ddagger}\right]=\delta_{S_1 S_2}, \quad\left[B_{k_1}, B_{k_2}^{\ddagger}\right]=2 \pi \delta\left(k_1-k_2\right).
\eeq

The perturbative expansion begins with the eigenstates of the free part of $H\p$.  These can be constructed as follows \cite{cahill76,mekink}.  The kink ground state $\vac_0$ is defined to be the state satisfying
\beq
\pi_0\vac_0=B_k\vac_0=B_S\vac_0=0. \label{v0}
\eeq
Similarly, an $n$ meson state $|k_1\cdots k_n\rangle_0$ is 
\beq
|k_1\cdots k_n\rangle_0=\Bd1\cdots\Bd n\vac_0.
\eeq

A basis of the kink sector is provided by states $\phi_0^m|k_1\cdots k_n\rangle_0$.  Any state in the kink sector may be decomposed in this basis.  In particular, if $|\psi\rangle$ is a Hamiltonian eigenstate in the kink sector, then we name the corresponding coefficients $\gamma$
\beq
|\psi\rangle=\sum_{m,n=0}^\infty |\psi\rangle^{mn}\hsp
|\psi\rangle^{mn}=\phi_0^m\ppink{n}\gamma_\psi^{mn}(k_1\cdots k_n)|k_1\cdots k_n\rangle_0.
 \label{gameqa}
\eeq
These coefficients can be found in a perturbative expansion.  Recalling that $Q_0^{-1/2}$ is of order $O(\sl)$, where $Q_0$ is the classical kink mass and $\sl$ is our perturbative parameter, this expansion can be written
\beq
\gamma^{mn}=\sum_i Q_0^{-i/2}\gamma_i^{mn}
\eeq
where $i$ is the order of the expansion.

In the present work, we are interested in Hamiltonian eigenstates $|k_1\rangle$ consisting of a single kink and a single meson of momentum $k_1$.  At leading order, kink-meson elastic scattering will be completely determined by the order $i=2$ perturbative correction to this state, which is determined by the coefficients $\gamma_{2k_1}^{mn}$.  In fact, the scattering amplitude only depends on a single coefficient, $\gamma_{2k_1}^{01}$. 

This coefficient was evaluated in Ref.~\cite{menorm}.  Let us recall its general form here.  First one separates out a $\delta$ function piece which depends on the choice of normalization
\beq
\gamma_{2k_1}^{21}(k_2)=\hat\gamma_{2k_1}^{21}(k_2)+2\pi\delta(k_2-k_1)Q_0\left(\hat\sigma_{ k_1}-\sigma_{ k_1}
\right) 
\eeq
where $\hat\gamma_{ k_1}(k_2)$ is continuous at $k_2= k_1$.  Then it can be written in terms of the functions $\rho$ and $\hat\gamma$, defined below, as
\beq
\gamma_{2 k_1}^{01}(k_2)=\frac{\rho_{ k_1}(k_2)-\hat \gamma_{2 k_1}^{21}(k_2)}{\ok 1-\ok{2}}. \label{g2}
\eeq

We have not yet defined the coefficients at $\ok 1=\ok 2$, corresponding to the location of the pole in Eq.~(\ref{g2}).   There are two such cases, corresponding to $k_1=\pm k_2$.  In the case $k_1=k_2$, the choice is physically irrelevant as it corresponds to a choice of normalization of the state.  In the case $k_1=-k_2$ the choice is physically relevant.  The states $|k_1\rangle$ and $|-k_1\rangle$ are degenerate Hamiltonian eigenstates, and so the choice of convention for treating this pole is equivalent to a choice of vector in this degenerate eigenspace.  Any choice will yield a Hamiltonian eigenstate $|k_1\rangle$, and so the choice must be made appropriately for the physics of a given problem.  This will be done in Sec.~\ref{calcsez}.

Finally, for completeness, let us recall the functions
\bea
\rho_{ k_1}(k_2)&=&
\frac{\lambda Q_0}{4\omega_{k_1}}V_{\I k_2 - k_1}
+\frac{\lambda Q_0}{8}\left(\frac{V_{\I  - k_1}}{\omega^2_{ k_1}}-\frac{\Delta_{- k_1 B}}{\omega_{ k_1}\sqrt{\lambda Q_0}}\right)V_{\I k_2}\label{rho}\\
&&+\sqrt{\lambda Q_0}\left[\ppinkp{2}\frac{\sqrt{\lambda Q_0}V_{- k_1 k\p_1 k\p_2}V_{-k\p_1-k\p_2k_2}}{16\omega_{ k_1}\okp1\okp2\left(\omega_{ k_1}-\okp1-\okp2\right)}\right.\nonumber\\
&&\left.+\ppin{k\p}\left(\frac{\left(-\okp{}\Delta_{k\p B}-\sqrt{\lambda Q_0}V_{\I  k\p}\right)
V_{-k\p- k_1 k_2}}{8\okp{}^2\omega_{ k_1}}+\frac{\sqrt{\lambda Q_0}V_{- k_1 k\p k_2}V_{\I -k\p}}{8\omega_{ k_1}\okp{}\left(\omega_{ k_1}-\okp{}-\ok2\right)}\right)\right.\nonumber\\
&&+\left.
\frac{ \left(-\ok2\Delta_{k_2 B}-\sqrt{\lambda Q_0}V_{\I  k_2}\right)V_{\I - k_1}}{8\omega_{ k_1}\ok2}\right]
-\frac{\lambda Q_0}{16}\ppinkp{2}\frac{V_{k_2k\p_1k\p_2}V_{- k_1-k\p_1-k\p_2}}{\omega_{ k_1}\okp1\okp2\left(\ok2+\okp1+\okp2\right)}
\nonumber
\eea
and
\bea
\hat\gamma_{2 k_1}^{21}(k_2)&=&\frac{3}{8}\left(-1+\frac{\ok2}{\omega_{ k_1}}\right)\Delta_{k_2 B}\Delta_{- k_1 B}-\frac{1}{4}\ppin{k\p}\left(\frac{\ok2}{\okp{}}+\frac{\okp{}}{\omega_{ k_1}}
\right)\Delta_{- k_1,-k\p}\Delta_{k_2k\p}\label{hg}\\
&&-\frac{\sqrt{\lambda Q_0}}{8\omega_{ k_1}}\left(\omega_{k_2}\Delta_{k_2 B}\frac{V_{\I - k_1}}{\omega_{ k_1}}+\omega_{ k_1}\Delta_{- k_1 B}\frac{V_{\I  k_2}}{\ok2}\right)+\frac{1}{8}\ppin{k\p}
\frac{\sqrt{\lambda Q_0}\Delta_{-k\p B}V_{- k_1 k_2 k\p}}{\omega_{ k_1}\left(\omega_{ k_1}-\ok2-\okp{}\right)}.\nonumber
\eea
Here we have defined the shorthand notation
\bea
\Delta_{k_1k_2}&=&\int dx \g_{k_1}(x) \g\p_{k_2}(x)\\
\I(x)&=&\pin{k}\frac{\left|{\g}_{k}(x)\right|^2-1}{2\omega_k}+\sum_S \frac{\left|{\g}_{S}(x)\right|^2}{2\omega_k}\nonumber\\
V_{k_1 k_2 k_3}&=&\int d x V^{(3)}(\sqrt{\lambda} f(x)) \mathfrak{g}_{k_1}(x) \mathfrak{g}_{k_2}(x) \mathfrak{g}_{k_3}(x)\nonumber\\
V_{\I k}&=&\int d x V^{(3)}(\sqrt{\lambda} f(x)) \I(x) \mathfrak{g}_{k}(x)\nonumber\\
V_{\I k_1 k_2}&=&\int d x V^{(4)}(\sqrt{\lambda} f(x)) \I(x) \mathfrak{g}_{k_1}(x) \mathfrak{g}_{k_2}(x).\nonumber
\eea
The indices $k_i$ here run over the zero mode $B$, all shape modes $S$ as well as the continuum modes $k$.  

\section{The Calculation} \label{calcsez}

In one-dimensional nonrelativistic quantum mechanics, there are two ways to calculate the reflection coefficient for a particle scattering off of a localized feature in the potential.  The first method is as follows.  One first solves the time-independent Schrodinger equation, imposing the boundary condition that there are no particles incoming from the right.  One next takes the inner product of the Hamiltonian eigenstate with an outgoing wave packet far to the left.  To get a nonzero answer, of course one must choose a Hamiltonian eigenstate whose energy is in the continuum.  Such states are not normalizable, but one may nonetheless define a sensible, finite inner product.

In the other method, one begins with an incoming wave packet on the left.  This is evolved in time using $e^{-iHt}$ until a late time, after it is far from the feature.  The part on each side will be a superposition of outgoing plane waves.  The coefficients of this decomposition into plane waves are the scattering amplitude as a function of momentum.

In the present note, we will consider the quantum field theory analogue of the first method.  In Ref.~\cite{ellong} we will redo the calculation using the second method, to check that the answers agree.

\subsection{Defining the Wave Packet}

Following the prescription above, we consider Hamiltonian eigenstates $|k_1\rangle$, consisting of a kink and a meson of momentum $k_1$.  How do we impose the boundary condition that there are no incoming mesons from the right?  Of course the Hamiltonian eigenstate itself has no dynamics, so the question itself only makes sense if we construct a wave packet of these Hamiltonian eigenstates.   A wave packet beginning largely near $x_0\ll -1/m$ with average momentum $k_0\gg 1/\sigma$ can be chosen to be
\beq
|t=0\rangle=\pin{k_1} e^{-\sigma^2(k_1-k_0)^2-i(k_1-k_0)x_0}|k_1\rangle.
\eeq
Recall that $|k_1\rangle$ is an eigenstate of the full kink Hamiltonian $H\p$, not just the free part, and so it is a sum of many different free eigenstates with various numbers of mesons.  It even includes the reflected part $|-k_1\rangle_0$ with some small coefficient of order $O(\lambda)$.  This small coefficient will be responsible for the scattering amplitude calculated here.

The wave packet is not a Hamiltonian eigenstate, so it evolves.  At time $t$ it is equal to
\beq
|t\rangle=\pin{k_1} e^{-\sigma^2(k_1-k_0)^2-i(k_1-k_0)x_0-iE_{k_1}t}|k_1\rangle. \label{t}
\eeq
Let us make the approximation $E_{k_1}=\ok 1$, which holds at leading order.  Now recall $\sigma\gg 1/m$.  This allows us to expand the frequency $\omega$
\beq
\ok{1}=\ok{0}+\frac{\partial \ok{0}}{\partial k_0}(k_1-k_0)+O\left((k_1-k_0)^2\right)=\ok{0}+\frac{k_0}{\ok 0}(k_1-k_0)+O\left((k_1-k_0)^2\right).
\eeq
The higher orders capture physics such as the spreading of the wave packet, which we will ignore from now on as they are small at large enough $\sigma$.  

Defining the position
\beq
x_t=x_0+\frac{k_0}{\ok 0} t
\eeq
one can rewrite (\ref{t}) as
\beq
|t\rangle=e^{-i\ok 0 t}\pin{k_1} e^{-\sigma^2(k_1-k_0)^2-i(k_1-k_0)x_t}|k_1\rangle. \label{t2}
\eeq
Apparently the main part of the wave packet is at position $x_t$ at time $t$.

Consider a pole in $|k_1\rangle$ at $k_1=-k_2$
\beq
|k_1\rangle=\pin{k_2}\left(F(k_1,k_2)+\frac{R(k_1)}{k_1+k_2}\right)|k_2\rangle_0 \label{ls}
\eeq
where $F(k_1,k_2)$ is analytic near $k_1=-k_2$.  At this point, we have not yet specified the function at the pole.  This choice, we have seen in earlier papers, corresponds to a choice of eigenstate $|k_1\rangle$ inside of a degenerate eigenspace.  Eq.~(\ref{ls}) is a Lippmann-Schwinger equation, and the ambiguity at the pole corresponds to the usual freedom to choose a solution in that context.

Inserting (\ref{ls}) into (\ref{t2}) one finds
\beq
|t\rangle=e^{-i\ok 0 t}\pin{k_2} I(k_2)|k_2\rangle_0\hsp
I(k_2)=\pin{k_1} e^{-\sigma^2(k_1-k_0)^2-i(k_1-k_0)x_t} \left(F(k_1,k_2)+\frac{R(k_1)}{k_1+k_2}\right).
\eeq
We see that the definition of the pole affects the integral $I(k_2)$.

\subsection{Choosing a Hamiltonian Eigenstate}

Consider a time $t$ such that $x_t\ll0$ and $\sigma\ll|x_t|$.  Physically the first condition means that the meson is not yet near the kink, while the second means that the meson wave packet is much smaller than the kink-meson separation.  This is not in contradiction with our standard assumptions that $\sigma\gg1/m$ and $\sigma\gg1/k_0$.   Choose an $r$ such that $\sigma\ll 1/r \ll |x_t|$.

Now let us evaluate $I(k_2)$ by integrating around a semicircle of radius $r$ that closes in the $+i$ side of the complex $k_1$ plane.  In principle, there may be contributions from nonanalytic parts of $F(k_1,k_2)$ far from $k_2=-k_1$.  From the general form of $|k\rangle$ found in other papers, this only occurs at real $k_2$ where it will contribute to other asymptotic final states.  We will simply ignore these, as our interest lies in elastic scattering. In Ref.~\cite{ellong} we will let $k_2$ be arbitrary and so will consider such processes.

Thus elastic scattering can only arise from the pole contribution
\beq
I(k_2)\Big|_{\rm{pole}}=\pin{k_1} e^{-\sigma^2(k_1-k_0)^2-i(k_1-k_0)x_t}\frac{R(k_1)}{k_1+k_2}.
\eeq
Recall that $\sigma r\ll 1$ and so the Gaussian factor is approximately unity.  Physically this is a consequence of the fact that the wave packet is so broad that there is negligible momentum smearing, although it is not so broad that the meson yet overlaps with the kink.

As $x_t\ll0$, the meson has not yet had time to scatter.  Thus, we wish to choose our initial condition such that no elastic scattering has occurred, so that $I(k_2)\big|_{\rm{pole}}$ vanishes.  As the residue $R(k_1)$ may in principle be nonzero, we achieve this by choosing the pole to be outside of our integration contour.  In other words, we wish to shift the pole in the $-i$ direction.  This can be accomplished with the choice
\beq
|k_1\rangle=\pin{k_2}\left(F(k_1,k_2)+\frac{R(k_1)}{k_1+k_2+i\epsilon}\right)|k_2\rangle_0.
\eeq
This $+i\epsilon$ is the same one that arises in the Lippmann-Schwinger equation for the ``in" state.

\subsection{Calculating the Scattering Probability}

Now that our initial eigenstate is determined, we may proceed to evolve it through the scattering, to $x_t\gg0$.  The appropriate contour for evaluating $I(k_2)$ closes in the $-i$ direction, and so picks up the pole at $k_1=-k_2-i\epsilon$.  The residue theorem then yields
\beq
I(k_2)\Big|_{\rm{pole}}=-iR(-k_2) e^{-\sigma^2(k_2+k_0)^2+i(k_2+k_0)x_t}
\eeq
and so the reflected part of the state is
\beq
|t\rangle_{\rm{reflected}}=-i e^{-i\ok 0 t}\pin{k_2} R(-k_2) e^{-\sigma^2(k_2+k_0)^2+i(k_2+k_0)x_t}|k_2\rangle_0.
\eeq
Here we have used the fact that, by energy conservation, the reflected part of the state is necessarily at $k_2=-k_1$ and so corresponds to the pole contribution.

As $\sigma |k_0|\gg1$ and $m\sigma\gg1$, one may approximate $R(-k_2)$ by its value at the peak of the Gaussian
\beq
|t\rangle_{\rm{reflected}}=-i e^{-i\ok 0 t} R(k_0)\pin{k_2} e^{-\sigma^2(k_2+k_0)^2+i(k_2+k_0)x_t}|k_2\rangle_0.
\eeq

The scattering probability is simply
\beq
P(k_0)=\frac{{}_{\rm{reflected}}\langle t|t\rangle_{\rm{reflected}}}{\langle t=0|t=0\rangle}=|R(k_0)|^2.
\eeq
The inner products were technically divergent as the states are nonrenormalizable.  Therefore they were computed using the reduced inner product of Ref.~\cite{menorm}.  This norm contains additional, subdominant terms, which change the meson number by one.  However it is clear that these vanish here, as all mesons, long after an interaction, are far from the kink but the subdominant terms contain factors of $\Delta_{kB}$ which have support close to the kink.  The only exception to this argument is Stokes scattering, in which the final state contains an excited shape mode, but then energy conservation would demand that the final state meson does not have momentum $-k_0$, which does not describe elastic scattering.  We will treat this term in Ref.~\cite{ellong} where we will see that it exactly cancels a final state correction.

\subsection{Calculating the Elastic Scattering Amplitude}

We have now seen that $R(k)$ is the elastic scattering amplitude for an incoming meson with momentum $k$.  The function $R(k)$ was, in turn, defined to be the residue of the pole in the Lippmann-Schwinger equation~(\ref{ls}).

Comparing the definition (\ref{ls}) with the result (\ref{g2}), one finds
\beq
R(k_0)=\frac{1}{Q_0}\frac{\partial k_ 0}{\partial \ok 0}\left( \rho_{ k_0}(-k_0) -\hat \gamma_{2 k_0}^{21}(-k_0)\right)=\frac{\ok 0}{Q_0 k_0}\left( \rho_{ k_0}(-k_0) -\hat \gamma_{2 k_0}^{21}(-k_0)\right).
\eeq
The expressions for $\rho$ and $\hat\gamma$ in Eqs.~(\ref{rho},\ref{hg}) simplify slightly to
\bea
\rho_{ k_0}(-k_0)&=&
\frac{\lambda Q_0}{4\ok{0}}V_{\I -k_0 - k_0}
-\frac{\sqrt{\lambda Q_0}}{4\ok 0}\Delta_{- k_0 B}V_{\I -k_0}\\
&&\hspace{-1.4cm}+\frac{\lambda Q_0}{16\ok 0}\ppinkp{2}\frac{V_{- k_0 k\p_1 k\p_2}V_{-k\p_1-k\p_2-k_0}}{\okp1\okp2}\left(\frac{1}{\ok 0-\okp1-\okp2+i\epsilon}-\frac{1}{\ok 0+\okp1+\okp2}\right)\nonumber\\
&&\hspace{-1.4cm}-\frac{\sqrt{\lambda Q_0}}{8\ok 0}\ppin{k\p}\frac{\left(\okp{}\Delta_{k\p B}+2\sqrt{\lambda Q_0}V_{\I  k\p}\right)
V_{-k\p- k_0 -k_0}}{\okp{}^2}\nonumber
\eea
and
\bea
-\hat\gamma_{2 k_0}^{21}(-k_0)&=&\frac{1}{4}\ppin{k\p}\left(\frac{\ok0}{\okp{}}+\frac{\okp{}}{\omega_{ k_0}}
\right)\Delta_{- k_0,-k\p}\Delta_{-k_0k\p}\\
&&+\frac{\sqrt{\lambda Q_0}}{4\omega_{ k_0}}\Delta_{-k_0 B}V_{\I - k_0}+\frac{\sqrt{\lambda Q_0}}{8\ok 0}\ppin{k\p}
\frac{\Delta_{-k\p B}V_{- k_0 -k_0 k\p}}{\okp{}}.\nonumber
\eea
Again we have chosen the sign in the pole to correspond the ``in" state from the Lippmann-Schwinger equation.  Note that the terms quadratic in $\Delta_{kB}$ have cancelled.  They would have led to elastic scattering with no intermediate mesons, corresponding to the Yukawa terms reported in Ref. \cite{uehara91}.

\begin{figure}[htbp]
\centering
\includegraphics[width = 0.45\textwidth]{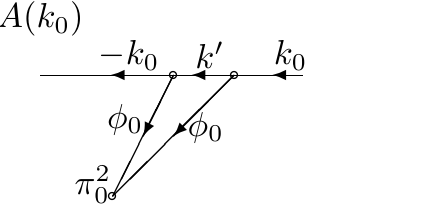}
\includegraphics[width = 0.45\textwidth]{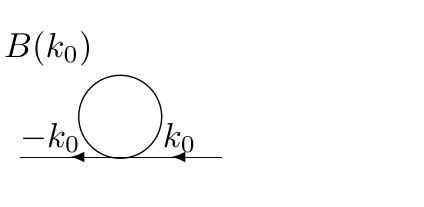}
\includegraphics[width = 0.45\textwidth]{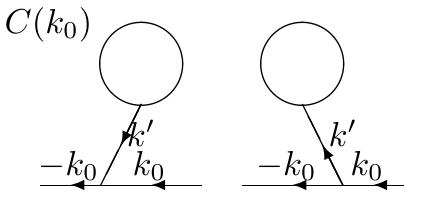}
\includegraphics[width = 0.45\textwidth]{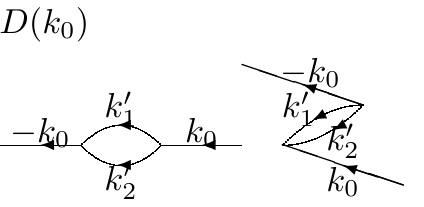}
\caption{Six diagrams represent the contributions $A(k_0)$, $B(k_0)$, $C(k_0)$ and $D(k_0)$ to the elastic scattering amplitude $R(k_0)$.  Here time runs to the left.}\label{abcdfig}
\end{figure}

Assembling these ingredients, one finds that the amplitude is the sum of the contributions from four kinds of processes
\beq
R(k_0)=\lambda(A(k_0)+B(k_0)+C(k_0)+D(k_0))
\eeq
where
\bea
A(k_0)&=&
\frac{1}{4\lambda Q_0 k_0}\ppin{k\p}\left(\frac{\ok0^2+\okp{}^2}{\okp{}}\right)\Delta_{- k_0,-k\p}\Delta_{-k_0k\p} \label{abcd}\\
B(k_0)&=&
\frac{V_{\I -k_0 - k_0}}{4 k_0}
\nonumber\\
C(k_0)&=&
-\frac{1}{4k_0}\ppin{k\p}\frac{V_{\I  k\p}
V_{-k\p- k_0 -k_0}}{\okp{}^2}
\nonumber\\
D(k_0)&=&
\frac{1}{8k_0}\ppinkp{2}\frac{\left(\okp1+\okp2\right)V_{- k_0 k\p_1 k\p_2}V_{-k_0 -k\p_1-k\p_2}}{\okp1\okp2\left(\ok 0^2-(\okp1+\okp2)^2+i\epsilon\right)}.
\nonumber
\eea
Correspondingly, the probability is the sum squared of these four contributions
\beq
P(k_0)=\lambda^2|A(k_0)+B(k_0)+C(k_0)+D(k_0)|^2. \label{peq}
\eeq

\subsection{The Four Processes}

The four kinds of processes are depicted schematically in Fig.~\ref{abcdfig}.  Only the meson lines are shown, although the kink is treated fully dynamically and one should remember that the internal lines $k\p$ have values which run over not only the continuum modes, but also any shape modes that the kink may possess.   $\I(x)$ is a loop factor that arises when a meson loop contains a single vertex.  The terms $V_{k_1\cdots k_n}$ are the coupling constants responsible for $n$-meson interactions.  On the other hand, $V_{\I k_1\cdots k_n}$ is the coupling constant for the interaction of $n$ mesons plus a meson loop.  The matrix $\Delta_{k_1 k_2}$ describes the interaction of a meson with the momentum of the kink center of mass, changing the meson momentum from $k_1$ to $k_2$.

In process $A$, an incoming meson scatters off the kink, giving a kick to the momentum of the kink's center of mass.  To see this, recall that $\phi_0$ and $\pi_0$ are the usual $x$ and $p$ in the quantum mechanical description of the kink center of mass, and the first collision multiplies the corresponding wave function by $x$.  Next the meson interacts again, yielding a $\phi_0^2$, while it bounces off the kink.  The free evolution of the kink contains a nonrelativistic kinetic term $\pi_0^2/2$ which eliminates the $\phi_0^2$ and returns the kink to its ground state after the meson has left.

The process $B$ corresponds to the meson reflecting off the kink, but the interaction is in fact a four-point interaction involving a virtual meson-antimeson pair that propagates briefly inside the kink.  

The process $C$ is similar, but when the meson reflects it leaves a virtual meson, which is annihilated by such a meson-antimeson pair.  In Ref.~\cite{metad} we showed that such tadpoles can be removed by a quantum correction to the classical kink solution appearing in the displacement function $\df$, chosen so as to include a linear term in the Hamiltonian which exactly cancels the tadpole diagram.  We saw that such a change of prescription does not affect the quantum corrections to the kink mass, it simply reorganizes them.  We suspect that also in the present case, such a choice would lead the tadpole diagrams to disappear, but the same contribution would arise from elsewhere. 

Process $D$ is the most interesting.  Here the meson reflects via the creation of two virtual mesons, one or both of which may be shape modes.  In particular, if they are both shape modes, we see that the denominator of $D(k_0)$ possesses a pole at which the incoming meson energy is the energy of the twice excited shape mode.  We expect that, including higher order terms, this pole will assume the usual Breit-Wigner shape of an unstable resonance, with a width equal to the inverse lifetime computed in Ref.~\cite{alberto}.  Note that there is no such contribution in Eq.~(3.19) of Ref.~\cite{uehara91}, in which there is at most one intermediate meson.

The denominator has an imaginary part, corresponding again to same $i\epsilon$ as in the Lippmann-Schwinger equation.  In this context, it was found in Ref.~\cite{memult} and an alternate derivation appeared in Ref.~\cite{menorm}.  The pole corresponds to the case in which the two-meson intermediate state is on-shell.  In the case of the Sine-Gordon model, the pole will be exactly canceled by a term in the $V_{- k_0 k\p_1 k\p_2}$ in the numerator, which is a consequence of the fact that meson multiplication is forbidden by integrability \cite{memult}.

\section{Example: The Sine-Gordon Model} \label{exsez}
The Sine-Gordon model is an integrable quantum field theory described by the potential
\beq
V(\sqrt{\lambda}\phi(x))=m^2\left(1-{\rm{cos}}(\sqrt{\lambda}\phi(x)\right).
\eeq
Its kink has profile
\beq
f(x)=\frac{4}{\sl}{\rm{arctan}}\left(e^{mx}\right)\hsp Q_0\lambda=8
\eeq
and is called the Sine-Gordon soliton, because it is a soliton in the original sense, it scatters without deformation.

In particular, as in all integrable field theories, the S-matrix only allows for elastic scattering of particles with the same mass.  The soliton and the meson do not have the same mass, and so their elastic scattering is not allowed.  It is therefore a critical test of our result that $R(k_0)=0$ in this case.

Let us now calculate $R(k_0)$.  From the potential and the soliton solution, one can easily find
\beq
V^{(3)}[\sqrt{\lambda} f(x)]=2 m^2  \tanh (m x)\sech (m x)\hsp
V^{(4)}[\sl f(x)]= m^2 \left(-1+2\sech^2(mx)\right).
\eeq
The normal modes $\g_k(x)$ of the Sine-Gordon kink, and the loop factor $\I(x)$ are given by
\beq
\g_k(x)=\frac{e^{-ikx}}{\ok{}}(k-im \tanh(mx))\hsp
\I(x)=-\frac{\sech^2(mx)}{2\pi}.
\eeq
From these, one can easily compute the other relevant functions
\bea
\Delta_{k_1k_2}&=&-i(k_1-k_2)\pi\delta(k_1+k_2)+\frac{i\pi}{2}\frac{(k_2^2-k_1^2)}{\omega_{k_1}\omega_{k_2}}{\rm{csch}}\left(\frac{\pi\left(k_1+k_2\right)}{2m}\right)\\
V_{\I-k-k}&=&\frac{k(4k^2+m^2)}{15m^2}\csch\left( \frac{k\pi}{m}\right)\hsp
V_{\I k}=\frac{i}{8m^2}\ok{}^3 \sech\left( \frac{k\pi}{2m}\right)\nonumber\\
V_{k_1k_2k_3}&=&\frac{\pi i(\ok 1+\ok 2+\ok 3)}{4\ok 1\ok 2\ok 3}(\ok 1+\ok 2-\ok 3)(\ok 1-\ok 2+\ok 3)\nonumber\\
&&\times(-\ok 1+\ok 2+\ok 3)
\sech\left( \frac{(k_1+k_2+k_3)\pi}{2m}\right).\nonumber
\eea

\begin{figure}[htbp]
\centering
\includegraphics[width = 0.65\textwidth]{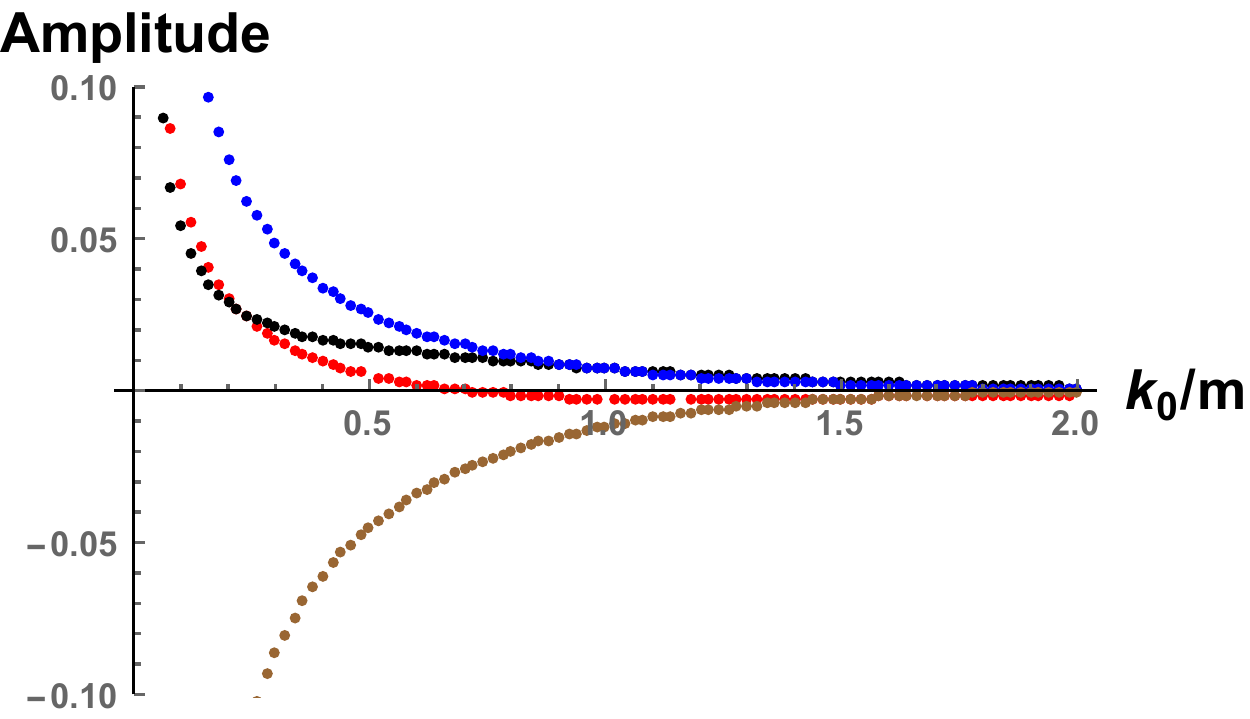}
\caption{Contributions $A$ (red), $B$ (black), $C$ (blue) and $D$ (brown) to the elastic scattering amplitude in the Sine-Gordon model}\label{sgfig}
\end{figure}

Substituting these into our general result (\ref{abcd}) one can find the individual contributions to soliton-meson scattering in the Sine-Gordon model
\bea
A(k_0)&=&\frac{\pi^2}{128m k_0\ok {0}^2}\pin{k\p}\frac{(\ok 0^4-\okp{}^4)(\okp{}^2-\ok 0^2)}{\okp{}^3}\csch\left( \frac{(k_0+k\p)\pi}{2m}\right)\csch\left( \frac{(k_0-k\p)\pi}{2m}\right)\nonumber
\\
B(k_0)&=&\frac{(4k_0^2+m^2)}{60m^2}\csch\left( \frac{k_0\pi}{m}\right)
\\
C(k_0)&=&\frac{\pi}{128m^2k_0\ok 0^2}\pin{k\p}\left(4\ok 0^2\okp {}^2-\okp{}^4\right)\sech\left( \frac{k\p\pi}{2m}\right)\sech\left( \frac{(2k_0+k\p)\pi}{2m}\right)
\nonumber\\
D(k_0)&=&-\frac{\pi^2 }{128 k_0\ok 0^2}\pinkp{2}\frac{(\okp 1+\okp 2)(\ok 0+\okp 1+\okp 2)^2}{\okp 1^3\okp 2^3
\left[(\okp 1+\okp2)^2-\ok 0^2 
\right]}(\ok 0+\okp 1-\okp2)^2\nonumber\\
&&\hspace{-1.5cm}\times(\ok 0-\okp 1+\okp2)^2 (-\ok 0+\okp 1+\okp2)^2\sech\left( \frac{(k\p_1+k\p_2+k_0)\pi}{2m}\right)\sech\left( \frac{(k\p_1+k\p_2-k_0)\pi}{2m}\right).\nonumber
\eea
These are each nonzero.  However, we have checked numerically that, up to at least one part in $10^{4}$, they cancel for all regimes of $k_0>0$.  The relative contributions can be seen in Fig.~\ref{sgfig}.

\section{Remarks}
This paper presented a quick and dirty calculation of the amplitude and probability of elastic kink-meson scattering.  It identified contributions from Lippmann-Schwinger pole terms, but it did not show that there are no other contributions.  Nonetheless, in the case of the Sine-Gordon model, it led to a seemingly miraculous cancellation of the amplitude which was demanded by integrability, and so provided a consistency check of our results.  In the future a more thorough investigation would nonetheless be warranted, perhaps by starting with an initial asymptotic meson state, evolving it, and calculating the probability that the final meson is moving backwards \cite{ellong}.

In the case of models whose kinks have shape modes, we have seen that our result contains a contribution $D(k_0)$ in which the intermediate state contains two excited shape modes.  These lead to poles in the scattering amplitude.  Near the poles, our perturbative expansion fails as the pole contribution is greater than $1/\sl$.  Here we should include bubble diagrams to fix this problem, and we suspect that after summing these bubble diagrams the pole will shift in the complex plane and assume the usual Breit-Wigner form for a resonance.

Is kink-meson scattering important?  Recently the interactions of kinks with radiation has entered the spotlight~\cite{mech22,multi22,dorey23,nav23,hahne23} as it has been discovered the interactions with bulk degrees of freedom play at least as important of a role in kink dynamics as shape modes~\cite{doreyprl}.  Yet, with some notable exceptions~\cite{kinkquantscat,alonso14,vach23}, so far this has largely been at the classical level, where reflectionless kinks are indeed reflectionless.  Here we see that at the quantum level, these interactions change qualitatively.

\section* {Acknowledgement}

\noindent
JE is supported by NSFC MianShang grants 11875296 and 11675223.

\end{document}